\documentclass[12pt]{article}
\usepackage{a4,psfig}
\usepackage{verbatim}
\usepackage{color}
\usepackage{rotate}
\usepackage{graphics}
\usepackage{epsf}
\usepackage{epsfig}
\usepackage{graphicx}
\usepackage{latexsym}
\usepackage{amscd}
\usepackage{amssymb}
\newcommand{\beq}{\begin{equation}}
\newcommand{\eeq}{\end{equation}}
\newcommand{\beqn}{\begin{eqnarray}}
\newcommand{\eeqn}{\end{eqnarray}}
\newcommand\noi{\noindent} 

\newcommand\ra{\rangle}
\newcommand\eps\varepsilon

\begin{document}

\hfill {LA-UR-00-5917}

\vspace*{3cm}

\centerline{\Large
\bf
The color dipole picture of the Drell-Yan process}
\vspace{.5cm}

\begin{center}
 {\large
B.Z.~Kopeliovich$^{a,d}$,
J.~Raufeisen$^{b}$
and
A.V.~Tarasov$^{c,d}$}\\
\medskip

{\sl
$^a$Max-Planck
Institut
f\"ur
Kernphysik,
Postfach
103980,
69029
Heidelberg,
Germany}\\

{\sl $^b$Physics Division,
Los Alamos National Laboratory, 
Los Alamos, New Mexico 87545}

{\sl $^c$Institut
f\"ur
Theoretische
Physik
der
Universit\"at,
Philosophenweg
19,
\\
69120
Heidelberg,
Germany}\\

{\sl
$^d$Joint
Institute
for
Nuclear
Research,
Dubna,
141980
Moscow
Region,
Russia}

\end{center}

\vspace{.5cm}

\begin{abstract}
 At high energies, Drell-Yan (DY)  dilepton production viewed in the target rest
frame should be interpreted as bremsstrahlung and can be expressed in terms of
the same color dipole cross section as DIS. We compute DY cross sections on a
nucleon target with the realistic parameterization for the dipole cross section
saturated at large separations. The results are compared to experimental data
and predictions for RHIC are presented.  The transverse momentum distribution of
the DY process is calculated and energy growth is expected to be steeper at
large than at small transverse momenta. We also calculate the DY angular
distribution and investigate deviations from the $1+\cos^2\theta$ shape.
\medskip

\noi
 \end{abstract}

\clearpage

\section{Introduction}\label{intro}

The Drell-Yan (DY) process in the kinematical region where the dilepton mass $M$ is
small compared to the center of mass energy $\sqrt{s}$ is of similar theoretical
interest as deep-inelastic scattering (DIS) at low Bjorken-$x$. Both processes probe the
target at high gluon density where one expects to find new physics. In contrast to DIS,
where only the total cross section can be measured, there is a variety of observables
which can be measured in the DY process, such as the transverse momentum distribution or
the angular distribution of the lepton pair.

The color dipole approach to the DY process suggested by one of the authors \cite{boris} 
(see also \cite{bhq}) provides a convenient alternative to the well known parton model, in
particular, it is especially appropriate to describe nuclear effects \cite{boris,kst}.
However, the dipole approach was not tested so far for the case of proton-proton
collisions. In this paper, we calculate different characteristics of the DY process on a
proton target.

\begin{figure}[ht]
  \scalebox{0.7}{\includegraphics*{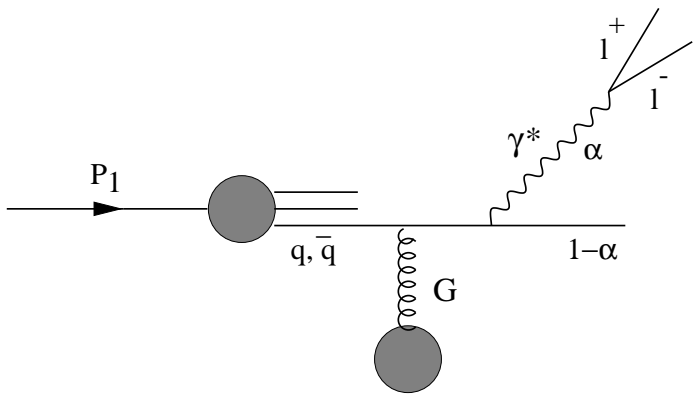}}\hfill
  \raise0cm\hbox{\parbox[b]{2.15in}{
    \caption{
      \label{bremsdy}
      In the target rest frame,
      DY dilepton production looks like bremsstrahlung. A quark 
      or an antiquark inside the
      projectile hadron scatters off the target color field and radiates a
      massive photon, which subsequently decays into the lepton pair. The photon
      can also be radiated before the quark hits the target. 
    }
  }
}
\end{figure}

Although cross sections are Lorentz invariant, the partonic interpretation of the
microscopic process depends on the reference frame.  In the target rest frame DY
dilepton production should be treated as bremsstrahlung, rather than parton
annihilation. The space-time picture of the DY process in the target rest frame is
illustrated in fig.\ \ref{bremsdy}.  A quark (or an antiquark) from the projectile
hadron radiates a virtual photon on impact on the target. The radiation can occur before
or after the quark scatters off the target. Only the latter case is shown in
fig.~\ref{bremsdy}.

A salient feature of the rest frame picture of DY dilepton production is that at high
energies and in impact parameter space the DY cross section can be formulated in terms
of the same dipole cross section as low-$x_{Bj}$ DIS.  Note that also the transverse
momentum distribution of DY dileptons can be expressed in terms of this dipole cross
section \cite{kst}.  The crucial input to all calculations is the dipole cross
section of interaction of a $q\bar q$ pair with a nucleon which at present cannot be
reliably calculated. We employ the parameterization of Golec-Biernat and W\"usthoff
\cite{Wuesthoff1}, since it describes well all DIS data in the range of $Q^2$ which is
relevant for DY.

As a result of the experimental situation, most work in low-$x$ physics is done for the
case of DIS where the structure function $F_2$ has been measured extensively at HERA. In
contrast to this, only very few data are available on the low-$x_2$ DY process. This
will however change with the advent of RHIC. In this paper, we perform the first
comparison between calculations in the dipole picture of DY and the available data.  In
section \ref{dipoledy} we give a description of the color dipole formulation of the DY
process. The results of the calculations for DY are presented in section
\ref{calculations}.

\section{The DY-process in impact parameter space}\label{dipoledy}

The cross section for radiation of a virtual photon from a quark after
scattering on a proton, can be written in factorized light-cone form
\cite{boris,bhq,kst}, 
\beq\label{dylctotal}
\frac{d\sigma(qp\to \gamma^*X)}{d\ln\alpha}
=\int d^2\rho\, |\Psi^{T,L}_{\gamma^* q}(\alpha,\rho)|^2
    \sigma_{q\bar q}(x_2,\alpha\rho),
\eeq
similar to the case of DIS.
Here, $\sigma_{q\bar q}$ is the cross section for scattering  a
$q\bar q$-dipole off a proton which depends on the $q\bar q$ separation 
$\alpha\rho$,
where $\rho$ is  the photon-quark transverse separation and $\alpha$ 
is the fraction of 
the light-cone momentum of the initial quark taken away by the photon.
We use the standard notation for the kinematical variables,
$x_2=(\sqrt{x_F^2+4\tau}-x_F)/2$, $\tau=M^2/s=x_1x_2$, where $x_F$ is the
Feynman variable,
$s$ is the center of mass energy squared of the colliding protons and 
$M$ is the
dilepton mass. In (\ref{dylctotal}) $T$ stands for transverse and $L$
for longitudinal photons.

An interesting feature of our approach is the appearance of the dipole 
cross section in
(\ref{dylctotal}), although there is no physical $q\bar q$-dipole in fig.\
\ref{bremsdy}.
The physical interpretation of (\ref{dylctotal}) is similar to the DIS
case. The projectile quark is expanded in the
interaction eigenstates. We keep only the first eigenstate,
\beq
|q\ra=\sqrt{Z_2}|q_{bare}\ra+\Psi^{T,L}_{\gamma^* q}|q\gamma^*\ra+\dots,
\eeq
where $Z_2$ is the wavefunction renormalization constant for fermions.
In order to produce a new state the interaction must resolve between the two Fock 
states, {\it i.e.} they have to interact differently. Since only the bare quarks
interact in both Fock components the difference arises from their relative displacement 
in transverse plane.
 If $\rho$ is the transverse separation between the
quark and the photon, the $\gamma^*q$ fluctuation has a center of gravity in the
transverse plane which coincides with the impact parameter of the parent quark.
The transverse separation between the photon and the center of gravity is
$(1-\alpha)\rho$ and the distance between the quark and the center of gravity is
correspondingly $\alpha\rho$. A displacement in coordinate space corresponds to
a phase factor in momentum space. The two graphs 
for bremsstrahlung, where the photon is radiated either 
before or after impact on the target,
 have the relative phase factor 
$-\exp({\rm i}\alpha\vec\rho\cdot\vec k_\perp)$, which produces the color 
screening
factor $[1-\exp({\rm i}\alpha\vec\rho\cdot\vec k_\perp)]$ in the dipole cross
section.

In Born approximation (two gluon exchange) 
the dipole cross section 
is independent of energy. 
The energy dependence  is generated by additional
radiation of gluons, which can be resummed in leading $\ln(1/x)$ approximation.
With help of the Weizs\"acker-Williams approximation
and at small separations, the dipole cross section can
be expressed in terms of the unintegrated target gluon density,
\beq
\label{borisdipole}
\sigma_{q\bar q}(x_2,\rho)=\frac{4\pi}{3}\alpha_s\rho^2
\int \frac{d^2 k_\perp}{k_\perp^2}
\frac{\left[1-\exp({\rm i}\vec k_\perp\cdot\vec\rho)\right]}
{k_\perp^2\rho^2}\,\frac{\partial\,G(x_2,k_\perp^2)}{\partial\,\ln(k_\perp^2)},
\eeq
where $k_\perp$ is the transverse momentum exchanged with the target.
This is explained in some detail in \cite{nzdipol}.
Note that the color screening
factor in (\ref{borisdipole}) makes the dipole cross section vanish like $\propto
\rho^2$ at $\rho\to 0$. This salient property of the dipole cross section 
is the heart
of the color transparency phenomenon \cite{zkl,bbgg,bm}.

In terms of Regge phenomenology, the color dipole approach accounts only for the
pomeron part of the cross section, since the dipole cross section
(\ref{borisdipole}) is governed by
gluonic exchange mechanisms. 
Therefore, this approach can be applied only at high
energies, i.e. at small $x_2$. 
As already mentioned above, the partonic interpretation of scattering
processes depend on the reference frame. In terms of the parton model, which is
formulated in the infinite momentum frame of the proton, the dipole approach
corresponds to annihilation of projectile
quarks (antiquarks) with sea antiquarks (quarks) of the target generated 
via gluons. 
Note that the statement, whether a sea quark belongs to the target or
to the projectile, is frame dependent.  If the projectile quark or antiquark
becomes slow in the limit $\alpha\to 1$, it can be interpreted in the infinite momentum
frame of the target as anti-seaquark or seaquark of the target which annihilates with
the projectile parton. No annihilation with valence quarks from the target is taken
into account in the dipole picture.  Note also 
that valence as well as sea parton
distributions of the projectile are contained in the parameterization of the projectile
structure function in (\ref{dylctotalhadr}).  Therefore, the formulation of the DY
process presented in this section is not fully symmetric between projectile and target.

In addition to sea quarks generated from gluon splitting 
there is a part of the sea generated nonperturbatively
from the meson cloud of the nucleon \cite{Sullivan}. 
This contribution has received much
attention in connection with the $\bar d/\bar u$
asymmetry measured recently by FNAL E866/NuSea \cite{NuSea}.
Since such a sea
component steeply decreases at small $x_2$, the dipole approach 
eq.~(\ref{dylctotal}) can be safely applied in this region.

The transverse momentum distribution of DY pairs
can also be expressed in terms of the dipole cross section \cite{kst}. 
The differential cross section is given by the 
Fourier integral
\beqn\nonumber\label{dylcdiff}
\frac{d\sigma(qp\to \gamma^*X)}{d\ln\alpha d^2q_\perp}
&=&\frac{1}{(2\pi)^2}
\int d^2\rho_1d^2\rho_2\, \exp[{\rm i}\vec q_\perp\cdot(\vec\rho_1-\vec\rho_2)]
\Psi^*_{\gamma^* q}(\alpha,\vec\rho_1)\Psi_{\gamma^* q}(\alpha,\vec\rho_2)\\
&\times&
\frac{1}{2}
\left\{\sigma_{q\bar q}(x_2,\alpha\rho_1)
+\sigma_{q\bar q}(x_2,\alpha\rho_2)
-\sigma_{q\bar q}(x_2,\alpha(\vec\rho_1-\vec\rho_2))\right\}.
\eeqn
after integrating this expression over the transverse momentum
$q_\perp$ of the photon, one obviously recovers
(\ref{dylctotal}). 
The expressions for the LC wavefunctions needed here are
\beqn\nonumber\label{dylct}
\Psi^{*T}_{\gamma^* q}(\alpha,\vec\rho_1)\Psi^T_{\gamma^* q}(\alpha,\vec\rho_2)
&=& \frac{\alpha_{em}}{2\pi^2}\Bigg\{
     m_f^2 \alpha^4 {\rm K}_0\left(\eta\rho_1\right)
     {\rm K}_0\left(\eta\rho_2\right)\\
   &+& \left[1+\left(1-\alpha\right)^2\right]\eta^2
   \frac{\vec\rho_1\cdot\vec\rho_2}{\rho_1\rho_2}
     {\rm K}_1\left(\eta\rho_1\right)
     {\rm K}_1\left(\eta\rho_2\right)\Bigg\},\\
\label{dylcl}
 \Psi^{*L}_{\gamma^* q}(\alpha,\vec\rho_1)\Psi^L_{\gamma^* q}(\alpha,\vec\rho_2)
&=& \frac{\alpha_{em}}{\pi^2}M^2 \left(1-\alpha\right)^2
  {\rm K}_0\left(\eta\rho_1\right)
     {\rm K}_0\left(\eta\rho_2\right), 
\eeqn
with $\eta^2=(1-\alpha)M^2-\alpha^2m_f^2$. We introduce a quark mass $m_f=200$
MeV. The quark mass has virtually no influence on the numerical results in $pp$
collisions, fig.\ \ref{dytotal}, but will be more important in proton-nucleus
collisions \cite{new}. 
Three of the four integrations in
(\ref{dylcdiff}) can be performed
analytically for arbitrary $\sigma_{q\bar q}$ \cite{thesis}.

For embedding the partonic cross section (\ref{dylctotal}) into the hadronic
environment, one has to note that the photon carries away the momentum fraction
$x_1$ from the projectile hadron. The hadronic cross
section reads then 
\beqn
\frac{d\sigma}{dM^2dx_F}&=&\frac{\alpha_{em}}{3\pi M^2}
\frac{x_1}{x_1+x_2}\int_{x_1}^1\frac{d\alpha}{\alpha^2}
\sum_fZ_f^2\left\{q_f\left(\frac{x_1}{\alpha}\right)+
q_{\bar f}\left(\frac{x_1}{\alpha}\right)\right\}
\frac{d\sigma(qp\to \gamma^*X)}{d\ln\alpha}\nonumber\\
\nonumber\\
&=&\frac{\alpha_{em}}{3\pi M^2}
\frac{1}{x_1+x_2}\int_{x_1}^1\frac{d\alpha}{\alpha}
F_2^p\left(\frac{x_1}{\alpha}\right)
\frac{d\sigma(qp\to \gamma^*X)}{d\ln\alpha},
\label{dylctotalhadr}
\eeqn
and similar for the transverse momentum distribution (\ref{dylcdiff}).
The factor $\alpha_{em}/(3\pi M^2)$ accounts for the decay of the photon
into the lepton pair.
Remarkably, the parton densities $q_f$, $q_{\bar f}$ 
of the projectile enters just in the combination
$F_2^p$, which is the structure function of the proton. 
Therefore we did not include the fractional quark charge
$Z_f$ in the DY wavefunctions (\ref{dylct}, \ref{dylcl}). 
The structure function
$F_2^p$ is needed at large values of $x_{Bj}$. We  employ the
parameterization from \cite{Mils} in our calculations.

The dipole cross section is largely unknown, only at small distances $\rho$ it can be
expressed in terms of the gluon density. However, several parameterizations exist in the
literature, describing the whole function $\sigma_{q\bar q}(x,\rho)$, without
explicitly taking into account the QCD evolution of the gluon density.  A very
economical parameterization is provided by the saturation model of Golec-Biernat and
W\"usthoff \cite{Wuesthoff1},
 \beq\label{wuestsigma}
\sigma_{q\bar q}(x,\rho)=
\sigma_0\left[1-\exp\left(-\frac{\rho^2Q_0^2}{4(x/x_0)^\lambda}\right)\right],
\eeq
where
$Q_0=1$ GeV and the three fitted parameters are 
$\sigma_0=23.03$ mb, $x_0=0.0003$, and $\lambda=0.288$.
This dipole cross section vanishes $\propto\rho^2$ at small distances, as
implied by color transparency and levels off exponentially at large separations,
which reminds one of eikonalization.
The authors of \cite{Wuesthoff1} are able to fit all available HERA data with 
a quite low $\chi^2$
and can furthermore also describe diffractive HERA data. Although the
parameterization (\ref{wuestsigma}) might be unrealistic at very large distances
(see discussion in \cite{KST2}),
we can safely use it, because DY data are all taken at quite large virtualities,
where (\ref{wuestsigma}) works well.

\section{Calculation of DY cross sections}\label{calculations}

\begin{figure}[t]
\centerline{
  \scalebox{0.43}{\includegraphics{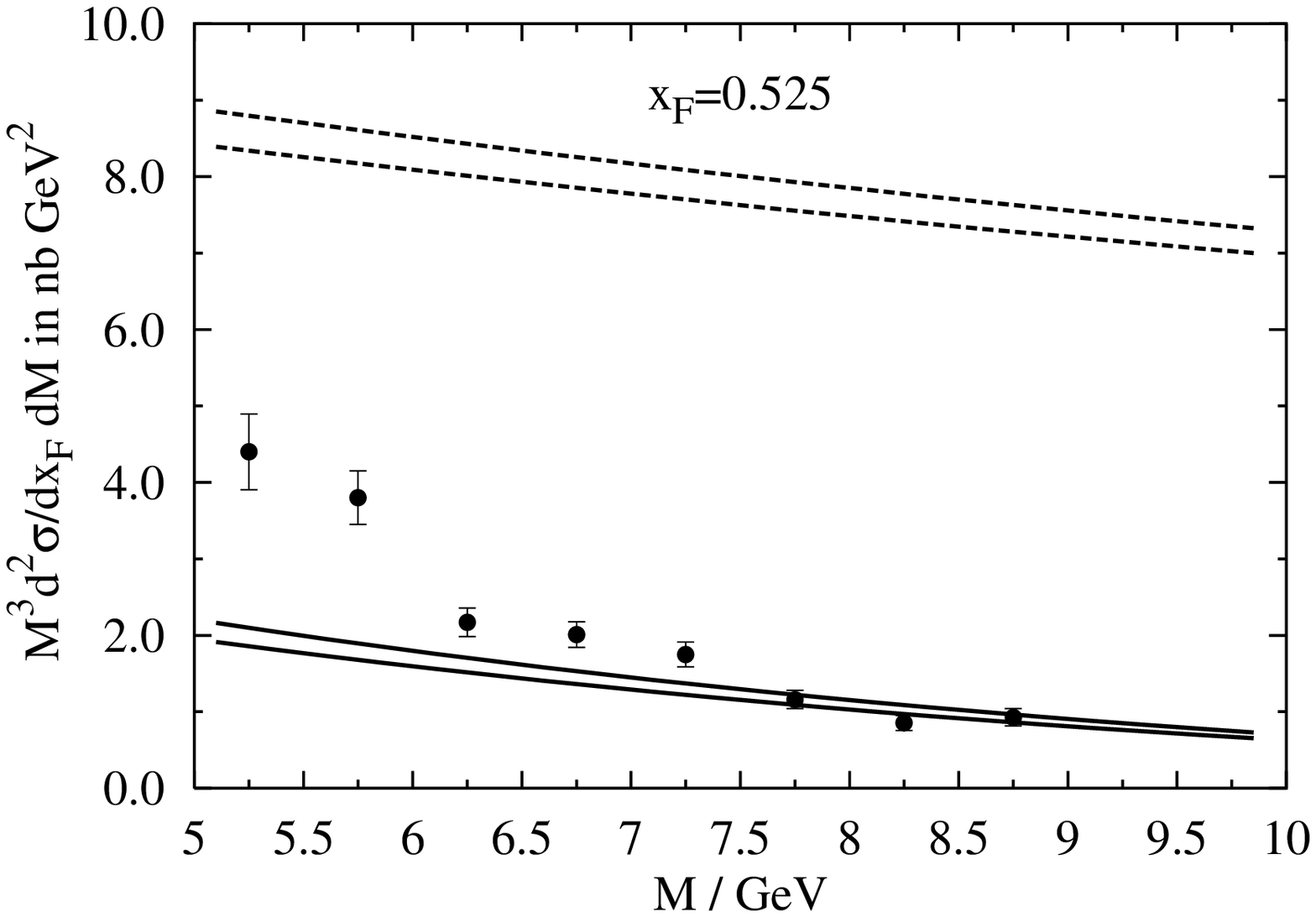}}
  \scalebox{0.43}{\includegraphics{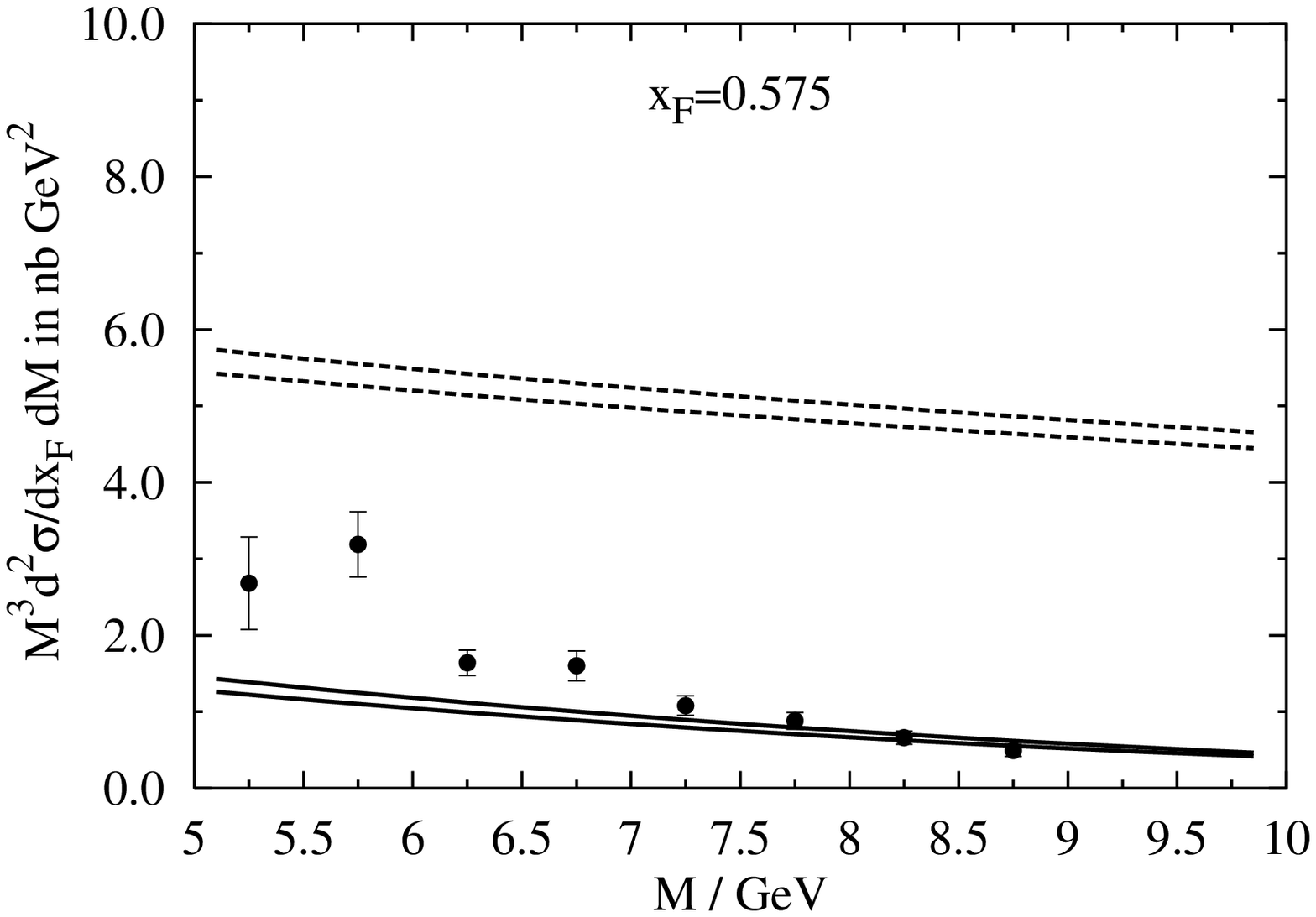}}
 }
\centerline{
  \scalebox{0.43}{\includegraphics{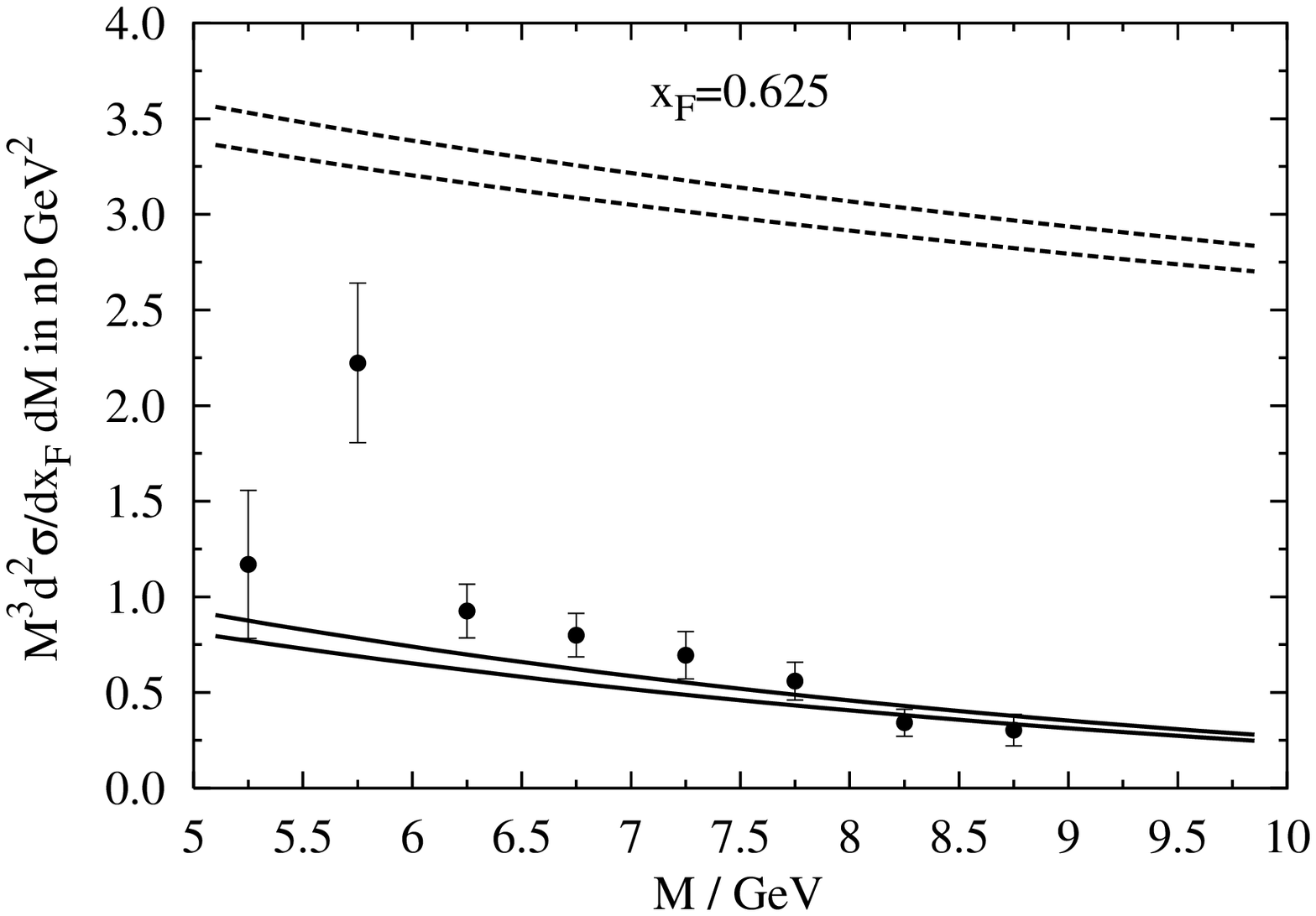}}
  \scalebox{0.43}{\includegraphics{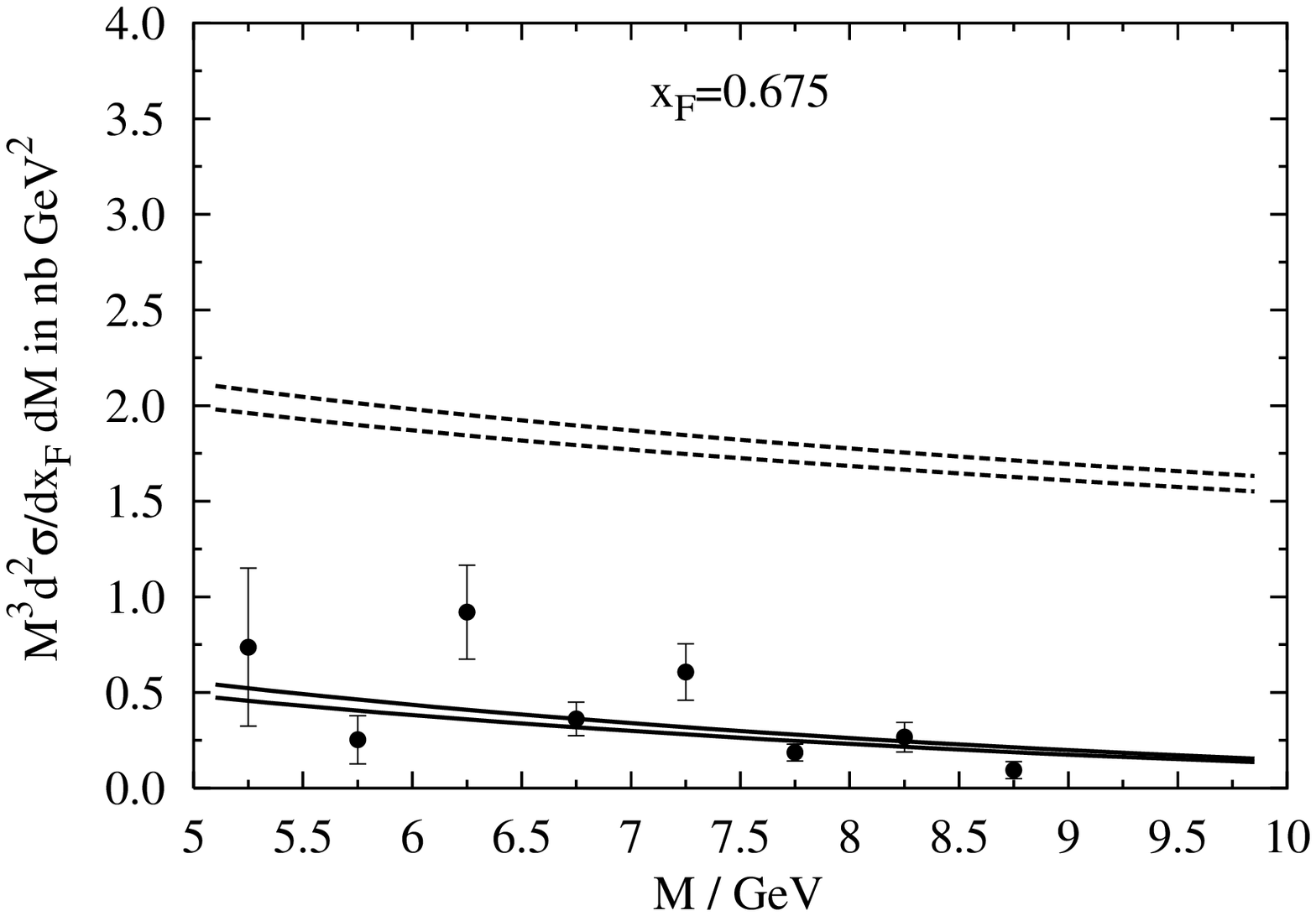}}
}
\caption{
  \label{dytotal}
  The points represent the measured DY cross section in $p\,^2H$ scattering
  from \cite{dydata}. Only
  statistical errors are shown. 
  Note that for these points $0.03\le x_2\le 0.09$. These values of $x_2$ are
  already quite large for the dipole approach.
  The curves are calculated with the dipole cross
  section (\ref{wuestsigma}) without any further
  fitting procedure.
  The solid curves are 
  calculated at the same kinematics as the data point (center
  of mass energy $\sqrt{s}=38.8$ GeV). The dashed curves are 
  calculated for RHIC
  energies, $\sqrt{s}=500$ GeV. 
  For each energy, the lower curve is
   for quark mass $m_f=200$ MeV, the upper curve for
  $m_f=0$.
}
\end{figure}

We can now proceed and investigate how well  DY
data are reproduced in the color dipole approach. 
At present, there are however not many
data for DY cross sections at low $x_2$. 
We compare to those data 
for $p\,^2\!H$ scattering
from E772 \cite{dydata} which correspond to the lowest
values of $x_2$. Since the dipole approach is valid at small $x_2$, we 
only compare
to points with $x_2 < 0.1$.
The result of
our calculation, using (\ref{dylctotalhadr}), 
is shown in fig.\ \ref{dytotal}. 
The curves for RHIC (dashed curves) correspond of course to 
lower values for $x_2$
than the ones for E772. The DY cross section increases, because the dipole cross
section increases with energy.
No further fitting procedure of the parameters in the dipole cross section 
(\ref{wuestsigma}) was
performed.
The data, and in particular the absolute
magnitude of the cross section is  
quite well reproduced, except for few points at low mass. We emphasize that the
curves in fig.\ \ref{dytotal} are results of a parameter free calculation. 
Varying the quark mass $m_f$ leaves the numerical results almost unaffected.
Note also that no
$K$-factor was introduced. 
As pointed out above, in pQCD
the dipole approach corresponds to a resummation of logarithms $\ln(1/x)$.
However, we do not perform a pQCD calculation, but employ the phenomenological
parametrization (\ref{wuestsigma}) which is fitted to DIS data. 
We assume that this parametrization contains also contributions beyond the
leading-log approximation, as well as nonperturbative effects.
Therefore, we believe that it is not legitimate to use a $K$-factor in
our approach.

\begin{figure}[t]
  \centerline{\scalebox{0.8}{\includegraphics*{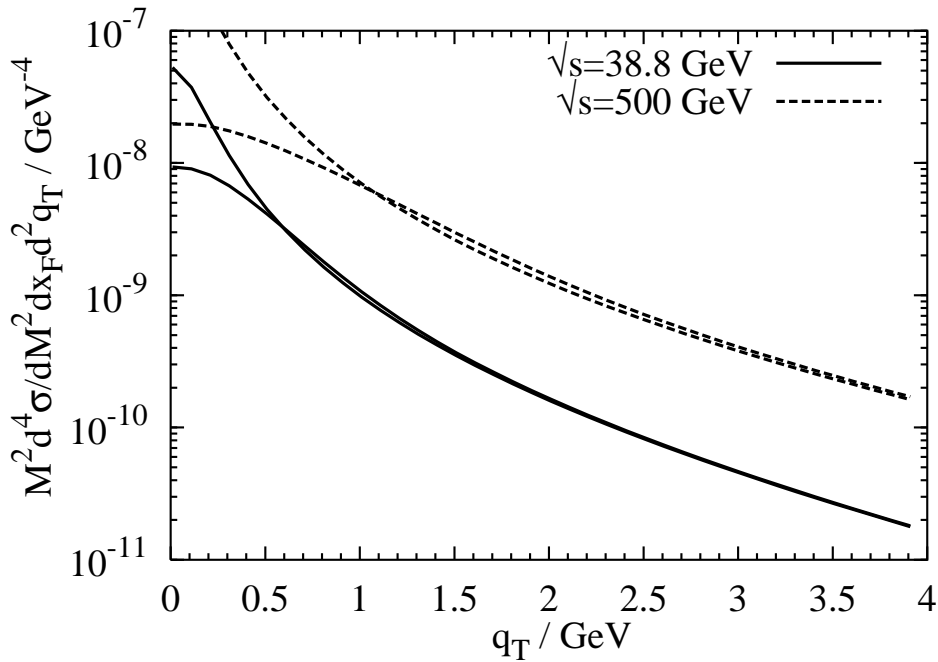}}}
    \caption{
      \label{dyperp}
      The transverse momentum distribution for DY pairs calculated from 
      (\ref{dylcdiff})  at
      $x_F=0.625$ and $M=6.5$ GeV. 
      The curves which flatten at small $q_\perp$ are calculated with
      the realistic dipole cross section
      (\ref{wuestsigma}), while the other two curves are calculated with the
      small $\rho$ approximation (\ref{sr}). 
    }  
\end{figure}

Furthermore, we also calculate the transverse momentum distribution of DY
dilepton pairs from (\ref{dylcdiff}). The result is depicted in fig.\
\ref{dyperp}. The DY cross section is finite at $q_\perp=0$, in contrast to
the first order pQCD correction to the parton model. In the parton model, one
has to resum large logarithms $\log(q_\perp/M)$ from soft gluon radiation
in order to avoid the divergence at $q_\perp=0$.  
In the dipole approach, the cross section does not diverge, because of the
saturation of the dipole cross section. In order to find out, how sensitive the
transverse momentum distribution to the large $\rho$-behavior of
$\sigma_{q\bar q}$ is, we do the same calculation with the small $\rho$
approximation of (\ref{wuestsigma}),
\beq\label{sr}
\widetilde\sigma_{q\bar q}(x,\rho)=\sigma_0\frac{Q_0^2}{4(x/x_0)^\lambda}\,\rho^2.
\eeq
The result is shown by the dotted curves in fig.\ \ref{dyperp}. There is no
divergence at $q_\perp=0$, because of the quark mass $m_f=200$ MeV, but the cross
section at small $q_\perp$ is quite strongly affected.

We do not compare to data in fig.\ \ref{dyperp}, because all available data are
integrated over $x_F$ and are therefore contaminated by valence quark
contributions. 
An extraction of the low-$x$ part of the transverse momentum distribution
is in progress \cite{private}. Note that the differential cross section
increases at large transverse momentum faster with energy than at low $q_\perp$. 
The
reason for this behavior is that at large $q_\perp$
small dipole
sizes are  predominantly sampled, where 
the dipole cross section increases more rapidly with
energy than at large separations. These large separations become more important
at low $q_\perp$. The resulting broadening of transverse momenta with energy
should not be
confused with the well known broadening at fixed $\tau=M^2/s$, which can be
understood from purely dimensional arguments.

\begin{figure}[t]
\centerline{
  \scalebox{0.43}{\includegraphics{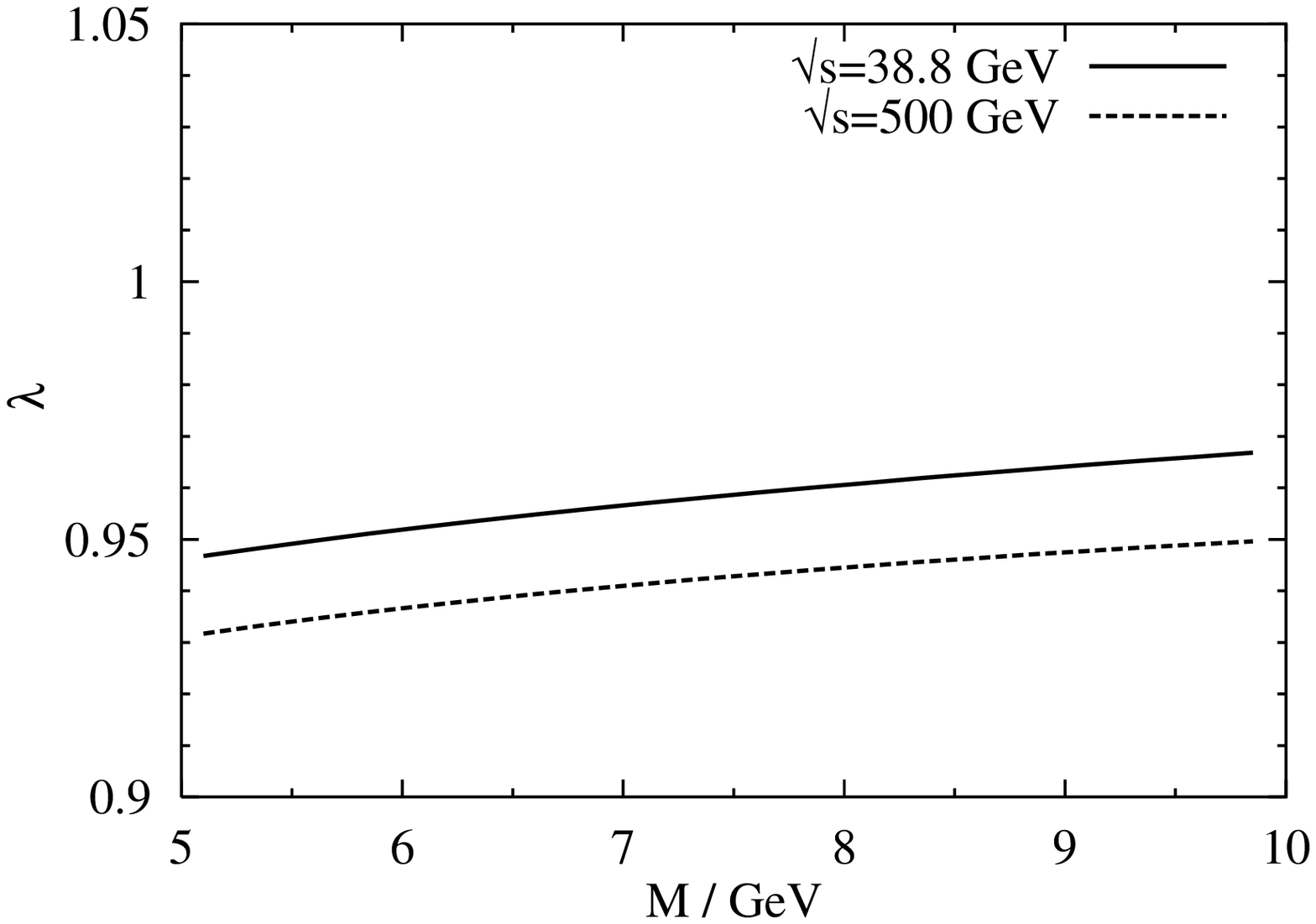}}
  \scalebox{0.43}{\includegraphics{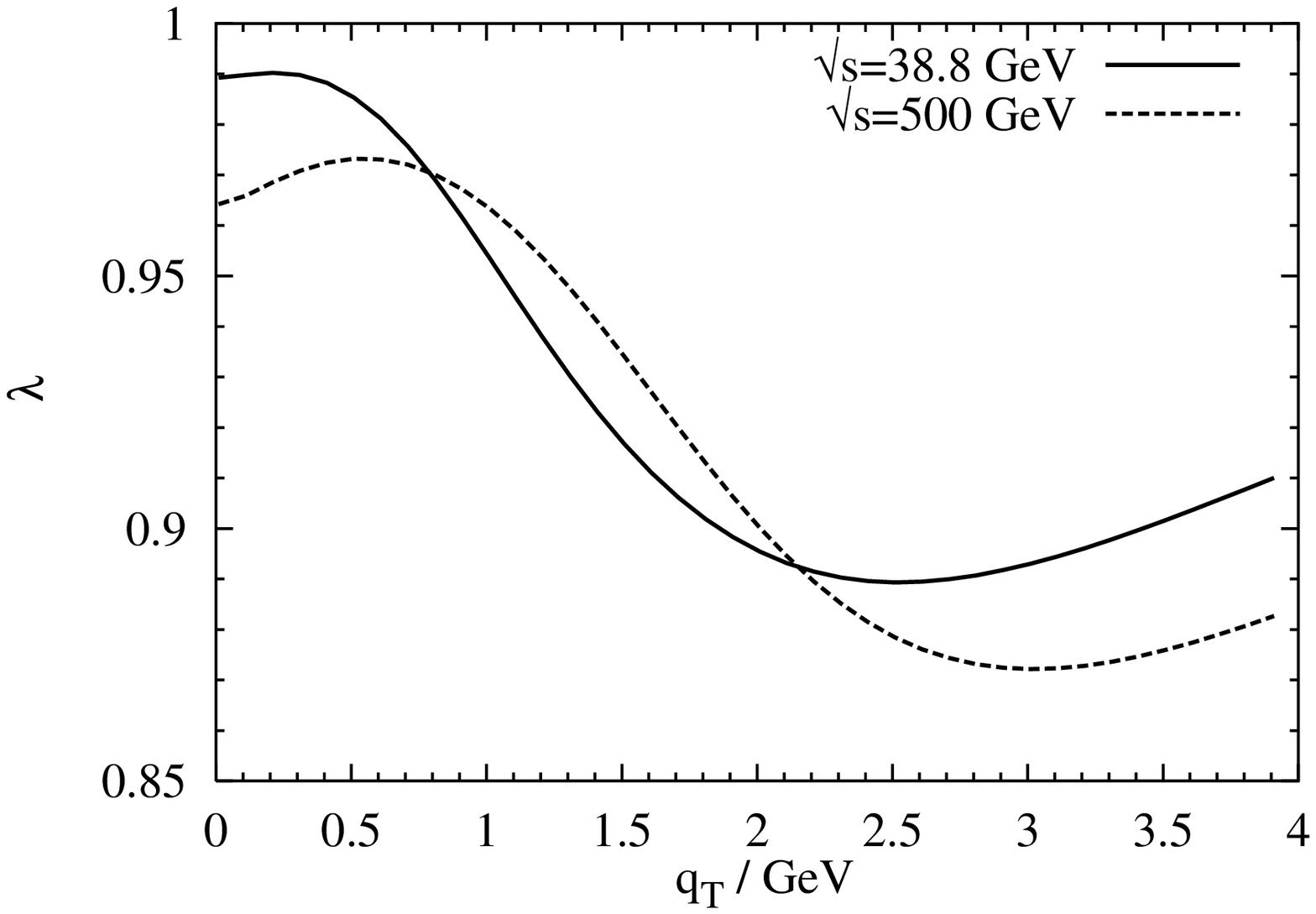}}
}
\caption{\label{lambpar}\label{lambparpt}
  The left figure shows the dependence of the
  parameter $\lambda$, which describes the angular
  distribution of DY pairs (\ref{angular}) on the dilepton mass.
  The calculation
  is performed for $x_F=0.625$.
  The figure on the right displays the $q_\perp$ dependence of $\lambda$ at
  $x_F=0.625$ and $M=6.5$ GeV.
}
\end{figure}

The color dipole approach allows one to calculate separately
the cross section for longitudinal
and transverse photons. Experimentally, different polarizations can be
distinguished by investigating the angular distribution of DY pairs, which 
can be written as
\beq\label{angular}
\frac{d\sigma}{dx_FdM^2d\cos\theta}\propto 1+\lambda\cos^2\theta,
\eeq
where $\theta$ is the angle between muon and the $z$-axis in the rest frame of
the virtual photon. The parameter $\lambda$ equals to $\pm1$
for transverse and longitudinal 
photons respectively.
Therefore, it can be calculated as
\beq
\lambda=\frac{\sigma_T-\sigma_L}{\sigma_T+\sigma_L}.
\eeq
Data for the angular distribution of DY pairs is usually presented 
in the dilepton center
of mass frame and the value of $\lambda$ depends on the choice of $z$-direction.
Since the dipole approach is formulated in the target rest frame, it is 
convenient to put the $z$-axis in direction of the radiated photon \cite{bhq}.
The target rest frame and the dilepton center of mass frame are then related 
by a boost in $z$-direction. Note that in the dilepton center of mass
frame, the $z$-axis is antiparallel to the target momentum. This frame is called
the $u$-channel frame and the curves we present for $\lambda$ are valid for this
frame.

We study the dependence of $\lambda$ on the dilepton mass and on the transverse
momentum of the pair.
Our results are shown in fig.\ \ref{lambpar}.
The deviation of $\lambda$ from unity decreases very slowly with increasing
mass. Although in fig.\ \ref{lambpar} (left) 
$\lambda$ is slightly smaller at RHIC energies than at E772 energies, the
deviation from a $1+\cos^2\theta$ distribution is typically a 5\% effect.
Note that in DY 
from pion-tungsten scattering at large $x_F$ a sudden change of the
angular distribution from $1+\cos^2\theta$ to $\sin^2\theta$ has been 
observed \cite{LTexp1}.
This is usually explained by interactions of the spectator quark \cite{BB}. 
Such mechanisms are not included in the dipole
approach. The K$_1$-part in the transverse light cone wavefunction always
dominates over the K$_0$ part in the longitudinal wavefunction. Thus, deviations
from the $1+\cos^2\theta$ shape are always small.

As function of the transverse momentum $q_\perp$, 
fig.\ \ref{lambparpt} (right), 
deviations from unity can become $\sim 10$\% and
$\lambda$
exhibits an interesting nonmonotonous behavior which can be checked in future
experiments.  
Note that in the parton model, the Lam-Tung relation \cite{LT} and helicity
conservation require that $\lambda(q_\perp=0)=1$, which is obviously not the case
in the dipole approach. Thus, the Lam-Tung relation is violated in the dipole
approach.
The reason for this
behavior is caused by nonperturbative effects, which are parameterized in the
dipole cross section. Using a parameterization $\propto\rho^2$ instead of
(\ref{wuestsigma}) 
would yield $\lambda(q_\perp=0)=1$, as can be seen from eq.~(16) in
\cite{kst}. 
Experimentally, the Lam-Tung relation is found to be violated
\cite{LTexp1,LTexp2}.

We did not calculate the $\phi$ dependence of the cross section, since there is
no hope that this will be measured within a foreseeable future.
The only way to check the Lam-Tung relation in the not too far future is to study
the limit $\lambda(q_\perp\to 0)$, which is possible at RHIC. It is a special
virtue of the dipole approach, that one can easily 
perform calculations at $q_\perp\ll
M$.

\section{Summary}

In this paper, the first realistic calculations in the color dipole approach to
the DY process in proton-proton collisions
are presented. We employ the parameterization \cite{Wuesthoff1} of
the dipole cross section and find good agreement with E772 data \cite{dydata} at
low $x_2$, without any $K$-factor or free parameter. 
The quark mass, which is in principle undetermined, has virtually no influence on
the numerical results.
The cross section steeply rises
with energy and is about four times larger at RHIC  than at Fermilab.

We also study the transverse momentum distribution of DY pairs. As a consequence of the
saturation of the dipole cross section at large separations, the differential
cross section does not diverge at zero transverse momentum, in contrast to the 
first order perturbative QCD correction to the parton model. The differential cross 
section rises with energy faster at large than at small momentum transfer.
This correlates with the fact that the dipole cross section rises with energy
steeper at small than at large separations.
 
We parameterize the angular distribution of the DY pairs as
$(1+\lambda\cos^2\theta)$ and calculate the coefficient $\lambda$ in the
$u$-channel frame as a function of dilepton mass and transverse momentum. 
We find that $\lambda$ as a function of $M$ 
is typically around $\sim 0.95$. In addition we find that
$\lambda$ does not go to unity for vanishing transverse momentum. This is a
consequence of the parameterization of the dipole cross section which we employ.
The behavior of $\lambda$ at small $q_\perp$ can be checked in proton-proton
collisions at RHIC and possibly could point out the presence of dynamics beyond
the QCD improved parton model.

\bigskip
{\bf Acknowlegdgments:}
This work was partially supported by the 
Gesellschaft f\"ur Schwer\-ionenforschung, GSI, grant HD H\"UF T,
by the European Network
{\em Hadronic Physics with Electromagnetic Probes,}
Contract No.~FMRX-CT96-0008, 
by the Gra\-du\-ier\-ten\-kol\-leg {\em Physikalische Systeme mit
vielen Freiheitsgraden} and by the U.S.~Department of Energy at Los Alamos
National Laboratory under Contract No.~W-7405-ENG-38. 
Part of this work was done, while
J.R. was employee of the Institut f\"ur Theoretische Physik der
Universit\"at Heidelberg. 

We are grateful to  J\"org H\"ufner 
and Mikkel Johnson for valuable discussions.

\end{document}